\pdfoutput=1
\documentclass[11pt]{article}

\usepackage{acl}

\usepackage{times}
\usepackage{latexsym}
\usepackage{enumitem}
\usepackage{graphicx}
\usepackage{booktabs}
\usepackage{multirow}
\usepackage{hyperref}
\usepackage{amssymb}
\usepackage{xspace}
\usepackage{comment}
\usepackage{etoolbox,caption,subcaption}
\usepackage{microtype}
\usepackage{colortbl}
\usepackage{bbding}
\usepackage{bm}
\usepackage{dsfont}
\usepackage{makecell}
\usepackage{listings}
\usepackage{amsmath}
\usepackage{color}
\usepackage[linesnumbered,ruled,vlined]{algorithm2e}
\title{Aligning Large Language Models for Controllable Recommendations}
\author{
    \begin{tabular}{c}
    Wensheng Lu$^1$ \quad Jianxun Lian$^2$ \quad Wei Zhang$^1$ \quad Guanghua Li$^1$ \\
    Mingyang Zhou$^1$ \quad Hao Liao$^{1*}$ \quad Xing Xie$^2$
    \end{tabular}
    \\
    \small
    \begin{tabular}{c}
    College of Computer Science and Software Engineering, Shenzhen University, China$^1$ \\
    Microsoft Research Asia$^2$ \\
    \end{tabular}
    \\
    \small
    \begin{tabular}{c}
    \texttt{2210273060@email.szu.edu.cn} \quad \texttt{jianxun.lian@outlook.com} \\
    \texttt{\{2210275010, 2210275050\}@email.szu.edu.cn} \\
    \texttt{\{zmy, haoliao\}@szu.edu.cn} \\
    \texttt{xingx@microsoft.com}
    \end{tabular}
}

\begin{document}

\maketitle
\begin{abstract}

Inspired by the exceptional general intelligence of Large Language Models (LLMs), researchers have begun to explore their application in pioneering the next generation of recommender systems — systems that are conversational, explainable, and controllable. However, existing literature primarily concentrates on integrating domain-specific knowledge into LLMs to enhance accuracy using a fixed task template, often overlooking the diversity of recommendation tasks and the ability of LLMs to follow recommendation-specific instructions. To address this gap, we first introduce a collection of supervised learning tasks, augmented with labels derived from a conventional recommender model, aimed at explicitly improving LLMs' proficiency in adhering to recommendation-specific instructions. Next, we propose a reinforcement learning-based alignment procedure to enhance LLMs' generalization ability.  Extensive experiments on two real-world datasets demonstrate that our approach significantly improves the capability of LLMs to respond to instructions within recommender systems, reducing formatting errors while maintaining a high level of accuracy.

\end{abstract}

\section{Introduction}

Recommender systems are designed to identify and suggest the most appropriate items to users from a vast array of candidates, based on the users' profiles, past interactions, and present intentions. Witnessing the impressive capabilities of Large Language Models (LLMs), such as knowledge retention, reasoning, and problem-solving, researchers are now exploring the integration of LLMs into the next wave of intelligent recommender systems, which aim to be conversational, explainable, and controllable. Bridging the gap between the general-purpose LLMs and the specific requirements of recommendation tasks poses a challenge. In this context, fine-tuning LLMs with domain-specific knowledge and recommendation-focused tasks emerges as a promising strategy to harness their potential for advanced recommendation purposes~\cite{bao2023tallrec,zhang2023recommendation,chen2023palr}. 
 
Typical approaches in the literature involve reformatting recommendation tasks — such as item reranking or click-through rate (CTR) prediction — into natural language constructs to facilitate the fine-tuning of LLMs. 
However, we have observed that LLMs fine-tuned through this straightforward method, albeit enhancing accuracy in offline evaluations, frequently generate outputs with domain-specific formatting errors. These errors may manifest as repeated items in the top-k recommendations or the inclusion of items previously interacted with by the user. Additionally, these LLMs exhibit a limited ability to adhere to diverse recommendation-specific instructions. This compromises their effectiveness as interactive agents in real-world recommender systems. A vivid example can be found in Section~\ref{sec:case}.


In this paper, we investigate the alignment of an LLM for recommender systems. Our objective extends beyond merely improving the recommendation accuracy of an original LLM; we aim to significantly enhance controllability and reduce formatting errors. Drawing inspiration from the Reinforcement Learning from Human Feedback (RLHF) framework~\cite{ouyang2022training}, our methodology is structured into two phases: the supervised learning (SL) stage and the reinforcement learning (RL) stage. To inject domain-specific knowledge and foster recommendation-relevant control capabilities within the LLM, we devise a series of fine-tuning tasks, including item recommendation, item search, category control, and category proportion control. These tasks often necessitate generating a list of items that not only meet users' instructions but also maintain high quality, despite the typically sparse ground-truth signals found in user behavior history. To tackle this issue, we propose augmenting supervised labels with predictions from a traditional recommender model, such as SASRec~\cite{kang2018self}. These augmented labels can help distill knowledge from the traditionally trained recommender model and meet the dynamic requirements of recommendation instructions.

After the SL stage, the LLM exhibits a significantly enhanced ability to follow recommendation-related instructions, surpassing existing approaches that solely fine-tune the LLM on item recommendation and search tasks. Nevertheless, the SL stage's data generation process inherently provides only positive examples for each instruction. To address scenarios where the LLM generates suboptimal responses, we introduce an RL stage with carefully crafted reward signals to further refine the LLM's capacity to follow instructions. To the best of our knowledge, this is the first study that employs both SL and RL stages to align LLMs for controllable recommendation purposes. We conduct comprehensive experiments on two real-world datasets, Steam and Amazon Movie, with results demonstrating that our method markedly improves the LLM's ability to follow instructions while simultaneously reducing formatting errors.
Our major contributions are summarized as follows:
\begin{itemize}[leftmargin=*]
    \item 
    We introduce a novel supervised learning stage, which encompasses a suite of tasks designed for enhancing controllability and label-augmentation by a teacher recommender model, to align an LLM into an interactive recommender agent.
    \item
    To mitigate formatting errors and improve the instruction-following generalization, We further design an alignment stage based on reinforcement learning with a variety of rewards that are tailored for the nuances of the controllable recommendation task.
    \item 
    Experiment results validate that our model markedly surpasses existing LLM-based recommendation models, and exhibits a robust capacity to follow users' instructions while maintaining a high level of recommendation precision. Source code is available at \url{https://github.com/microsoft/RecAI/tree/main/RecLM-gen}.
\end{itemize}

\section{Methodology}
\subsection{Intention Categories}\label{sec:intent_category}
This paper aims to enhance the instruction-following capabilities of LLMs for recommendation tasks. We categorize recommendation instructions into three distinct types:

\noindent \textbf{Implicit intention}. This is the assumed default where the prompt describes the user's profile (such as attributes and past favored items). The LLM is tasked with recommending items that align with the user's preferences.

\noindent \textbf{Item-wise intention}. In addition to the profile, users may express specific desires, such that the recommended items should either exhibit particular characteristics (``I wish to watch an action movie'') or exclude them (``Please avoid suggesting any action movies'').

\noindent \textbf{List-wise intention}. Users may have requirements for the entire list of recommended items; hence, evaluating individual items' attributes is insufficient. For example, if all items in a recommendation list belong to the same category A, the user may be disappointed by the lack of diversity. Consequently, the user might request the recommender system to ensure that the proportion of category A is below a certain threshold.

\begin{figure*}
\setlength{\abovecaptionskip}{0.1cm}
\centering
\includegraphics[width=1.0\linewidth]{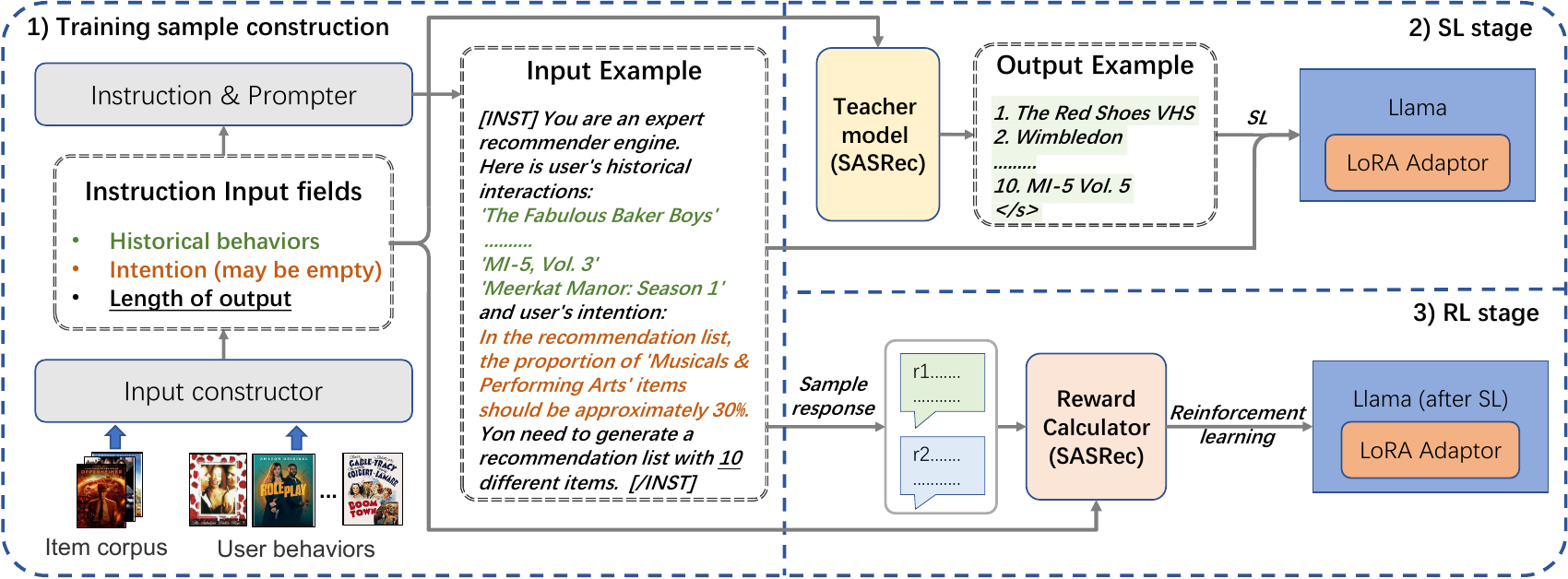}
\caption{Overview of the proposed method.}
\label{fig:framework}
\vspace{-0.3cm}
\end{figure*}

To effectively train LLMs as recommender agents capable of adhering to these three types of instructions, we introduce a novel two-stage fine-tuning approach: a supervised learning (SL) stage (Section \ref{sec:sl}) followed by a reinforcement learning (RL) stage (Section \ref{sec:rl}). The overall framework is illustrated in \autoref{fig:framework}.

\subsection{The SL Stage}\label{sec:sl}
We represent each data sample in the recommendation task with natural language text, adopting the format ``Instruction: [Prompt Content]. Output: [Response Content]'', where [Prompt Content] includes detailed instructions such as the user's profile and intention, and [Response Content] contains the expected item recommendations that fulfill the instructions. Different from traditional recommender systems that utilize item IDs, we employ only item titles to represent items in both [Prompt Content] and [Response Content] to fully leverage LLMs' general abilities and ensure a smooth interaction between users and our LLM-based recommender. 
Data samples are generated according to the following tasks:

\subsubsection{Data Generation Tasks}

\paragraph{Sequential Recommendation Instructions (${I_0}$)}
This task represents the traditional sequential recommendations: given a user's previously interacted items, the goal is to predict future interactions. Specifically, we use the first $n-1$ items to construct the user's behavioral profile, while the $n^{th}$ item serves as the ground-truth label. The LLM is instructed to recommend $k$ items; if the ground-truth item is among these $k$ suggestions, the recommendation is considered a successful hit. The value of $k$ for each data sample is randomly selected from $1$ to $10$.

\paragraph{Category Control Instructions (${I_1}$)}
This task corresponds to instructions with item-wise intention, which we implement into two distinct types: 1) Positive control (${I_{1}^{+C}}$), where the user hopes to receive more items in the recommendation list that match the specific category $C_{target}$.  2) Negative control (${I_{1}^{-C}}$), where users indicate a weariness towards a certain category $C_{target}$ and wish to reduce the inclusion of such items in their received recommendations as much as possible.

\paragraph{Category Proportion Control Instructions (${I_2}$)}  
We implement the list-wise intention into three distinct types: 1) ${I_{2}^{CP\leq x}}$, in this case, the user hopes to have the proportion of the item of $C_{target}$ less than a certain percentage $x$. 2) ${I_{2}^{CP\approx x}}$, in this case, the user hopes that the proportion of items in $C_{target}$ will be approximately a certain percentage. 3) ${I_{2}^{CP\geq x}}$, in this case, the user hopes that the proportion of items in $C_{target}$ is greater than a certain percentage.

\paragraph{Item Search Instructions (${I_3}$)}
To aid the LLM in memorizing in-domain item attributes (in this paper, we use item category as the key attribute for illustration), we introduce an item search task: the objective is to retrieve $k$ items belonging to a target category $C_{target}$. For this purpose, we randomly select $k$ items from $C_{target}$ to serve as the ground truth for the response.

\paragraph{ShareGPT (${I_4}$)}
To avoid catastrophic forgetting and preserve the general intelligence capabilities of the LLM, we follow \citet{zeng2023agenttuning} and integrate a certain proportion of ShareGPT\footnote{\url{https://huggingface.co/datasets/anon8231489123/ShareGPT_Vicuna_unfiltered}} training data into the SL stage. The ShareGPT data includes a diversity of real-world tasks in the user-assistant conversation format, which helps LLM revisit past knowledge during the in-domain fine-tuning stage.

\subsubsection{Label Augmentation}
Instructions ${I_0}$, ${I_1}$, and ${I_2}$ require the inclusion of $k$ items within the response text. However, due to the typically sparse nature of user historical behavior, it is often impractical to construct a ground-truth response based solely on this information. To overcome this limitation, we employ the sequential recommender model SASRec~\cite{kang2018self} as a teacher model to generate a set of top recommendations, $\bm{P}_{SASRec}$, for each data sample. We then curate the top-$k$ list by selecting items from $\bm{P}_{SASRec}$ that align with the given instructions. The top-$k$ list is assembled as follows: the first item is the ground-truth item (i.e., the $n^{th}$ item in the user's history). The method for compiling the remaining $k-1$ items varies slightly: for ${I_0}$, we fill the list with SASRec's top predictions; for ${I_1}$ and ${I_2}$, we filter SASRec's predictions to ensure the final recommendation list adhere to the specified instruction.

\subsection{The RL Stage}\label{sec:rl}
Following previous works~\cite{touvron2023llamachat,ouyang2022training}, we employ the Proximal Policy Optimization (PPO) algorithm~\cite{schulman2017proximal} to further fine-tune the model after the SL stage. 
Different from \cite{ouyang2022training}, scores are not produced by a reward model; instead, they are derived from reward rules that are specifically tailored for ${I_0}$, ${I_1}$, and ${I_2}$. Fundamentally, rewards consist of two components: item-level rewards and list-level rewards.

\subsubsection{Item-level Reward}
\label{sssec:3_3_2}
The item-level reward assigns a score to each item generated by LLM, serving as an immediate reward of reinforcement.

For each $item_i$ in LLM's recommendation list, let $Rank_i$ refer to the rank of $item_i$ in the teacher model (SASRec)'s prediction list. We can then calculate the preference scores for $item_i$ by:
\begin{equation}
\label{eq:score_p}
\resizebox{0.43\textwidth}{!}{
$Scores_i =
\begin{cases}
-1,  & \text{if $item_i$ is illegal} \\
+1, & \text{if $item_i$ is $item_{target}$} \\
\frac{1}{log_2(Rank_i+3)}, & \text{else}
\end{cases}$
}
\end{equation}
$item_i$ is considered illegal if it meets any of the following conditions: $item_i$ does not exist, it is a duplicate of any item within the preceding set $item_{[1,...,i-1]}$, it is identical to an item in the user's history, or its index $i$ exceeds $k$.

Beyond the preference score, $Scores$, a control effect score, $Scores^{ctl}$, is required to gauge the extent to which an item corresponds with user intentions. Essentially, a generated item that adheres to the given instruction is awarded a high score, signifying positive reinforcement, while a non-conforming item incurs a low score, serving as negative feedback. The complete calculation is described in \autoref{alg:item_wise_reward} in the Appendix.

Finally, we get item-level reward $R_{item}$, which measures the overall merits of each item:
\begin{equation}
\label{eq:reward_item}
R_{item} = 0.5*Scores + 0.5*Scores^{ctl}
\end{equation}

\subsubsection{List-level Reward}
The list-wise reward, which measures how well the entire list of recommendations matches the user's preferences and control intentions, is added to the last token of output, serving as a termination reward.
To encourage the LLM to rank ground-truth items in top positions, we adjust the $Scores$ as per \autoref{eq:score_p_correct} to derive $Score^*$:
\begin{equation}
\label{eq:score_p_correct}
Scores_i^* =
\begin{cases}
-1,  & \text{if $item_i$ is illegal} \\
\frac{Scores_i}{log_2(i+2)}, & \text{else}
\end{cases}
\end{equation}

In addition to this, we need to measure how well the output matches the control intention. Let $Count_{in},Count_{out}$ denote the number of items that belong/do not belong to the target category. For different control intentions, we use different calculation methods to get $Score^{ctl}_{list}$:
\begin{equation}
\label{eq:score_ctl}
\resizebox{0.43\textwidth}{!}{
$Score^{ctl}_{list} =
\begin{cases}
\mathrm{sum}(Scores^*),  & \text{if $I_0$} \\
\frac{1}{log_2((k-Count_{in})+2)},  & \text{if $I_1^{+C}$} \\
\frac{1}{log_2((k-Count_{out})+2)},  & \text{if $I_1^{-C}$} \\
\frac{1}{log_2(max(Count_{in}-k*m,0)+2)},  & \text{if $I_2^{CP\leq m}$} \\
\frac{1}{log_2(max(k*m-Count_{in},0)+2)},  & \text{if $I_2^{CP\geq m}$} \\
\frac{1}{log_2(abs(Count_{in}-k*m)+2)},  & \text{if $I_2^{CP\approx m}$} \\
\end{cases}$
}
\end{equation}
Finally, we get $R_{list}$, which measures the overall merits of the recommendation list:
\begin{equation}
\label{eq:reward_list}
\resizebox{0.43\textwidth}{!}{
$R_{list} = 0.5*\mathrm{sum}(Scores^*) + 0.5*Score^{ctl}_{list}$
}
\end{equation}

\subsubsection{RL Implementation Notes}
We follow the \textsl{Transformer Reinforcement Learning}\footnote{\url{https://github.com/huggingface/trl}} package to implement the reinforcement learning stage. We set up LoRA layers~\cite{hu2021lora} for both the policy network and the critic network, with a LoRA rank of 4 and a LoRA alpha of 2. During the RL sampling phase, we sample 2 responses with a temperature of $0.7$ for each instruction. The final reward is calculated by combining item-level reward, list-level reward, and a KL penalty:
\begin{equation}
\bm{R}_{\mathrm{final}}[\bm{y}]=\bm{R}_{item}[\bm{y}] + \bm{R}_{list}[\bm{y}] -\eta\bm{\mathrm{KL}}(\pi_{\theta}^{\mathrm{RL}},\pi^{\mathrm{SFT}})[\bm{y}]
\end{equation}
where $\eta$ is set to $0.3$. $\bm{y}$ represents a generated response (which is a token sequence) for an instruction sample.  $\bm{\mathrm{KL}}[\bm{y}]$ represents there is a KL penalty on each token in $\bm{y}$. $\bm{R}_{item}[\bm{y}]$ means there is an item-level reward at the last token of each item title. $\bm{R}_{list}[\bm{y}]$ represents that there is a list-level reward on the ending token of $\bm{y}$. To ensure that the information in $R_{list}$ is not overshadowed by $R_{item}$, we amplify $R_{list}$ by a factor of 10. Additionally, we employ reward whitening to enhance the stability of training.
We use Generalized Advantage Estimation (GAE) to estimate the advantage values, with hyperparameters $ \gamma=0.99, \lambda=0.95$. During the loss calculation phase, we set the clipping range for the probability ratio to $[1-\epsilon, 1+\epsilon]$, where $\epsilon=0.2$. The loss weight for the critic network is $0.5$. 

\section{Experiments} 
\subsection{Experiment Setting}
\subsubsection{Dataset} 

We experiment with two popular datasets in the recommender system domain: the Amazon Movies \& TVs~\footnote{\url{https://cseweb.ucsd.edu/~jmcauley/datasets/amazon_v2/}} dataset (\textbf{Movie} for short) and the Steam dataset~\cite{kang2018self}. Both datasets include item category information, which aids in constructing the user's control intentions. 
Basic statistics of datasets are summarized in \autoref{tab:DatasetStatistic}. We employ the leave-one-out approach~\cite{kang2018self} to split the dataset. Therefore, the size of test set is consistent with the number of users in each dataset. To accelerate the validation process, we only include four types of instructions in the valid set: $I_0$, $I_1^{+C}$, $I_1^{-C}$, and $I_2^{CP\approx50\%}$, and each type of instruction has 320 instances. Upon observing multiple metrics on the validation set, we find that SL typically converged around 30 epochs. Full details of the training set on each instruction task can be found at \autoref{tab:DatasetDetail} in the Appendix.
\begin{table}
\setlength{\abovecaptionskip}{0.1cm}
\resizebox{0.47\textwidth}{!}{
\begin{tabular}{l|cccc}
\toprule
$Dataset$ & $\mathrm{\#Users}$ & $\mathrm{\#Items}$ & $\mathrm{\#Inters}$ & $\mathrm{\#Sparsity}$ \\
\midrule
Movie  & $13,218$ & $18,744$ & $744,313$ & $99.70\%$ \\
Steam   & $12,658$ & $8,572$ & $632,900$ & $99.42\%$ \\
\bottomrule
\end{tabular}
}
\caption{General Statistics of the Two Datasets}
\label{tab:DatasetStatistic}
\vspace{-0.3cm}
\end{table}

\subsubsection{Implementation Details}


We choose Llama-2-7b-chat as the foundational model for our research \cite{touvron2023llamachat}. We set the model's maximum sequence length to 1024 tokens. User behavior sequences are truncated to incorporate no more than 10 items, and excessively lengthy item titles are condensed to a maximum of 64 tokens. Furthermore, to accommodate the complete recommendation list within the output, we reserve a larger token count, setting the maximum output length of our model to 512 tokens. Instructions and responses are formatted using the official prompt template in \citet{touvron2023llamachat}.


We use LoRA~\cite{hu2021lora} to fine-tune all linear layers in Llama2-7b-chat, with trainable parameters accounting for about 0.6\% of the total parameters in SL and 0.3\% in RL. The optimizer is Adam. In the SL stage, the learning rate is set to $0.001$, the LoRA dimension is set to 16, the LoRA alpha to 8, and the batch size is 64. The ShareGPT data accounted for 50\% of the total training data. In the RL stage, the learning rate is $5\times10^{-6}$. We set separate LoRA layers for actor and critic networks, with LoRA dimension at $4$ and LoRA alpha at $2$. To encourage model exploration, we set the temperature to $0.7$ and the weight of entropy loss to 0.01. 
We release the source code at: \url{https://github.com/microsoft/RecAI/tree/main/RecLM-gen}.

\subsubsection{Metrics}
For all models, we use Top-k hit ratio (\textbf{HR@K}) and Top-k NDCG (\textbf{NDCG@K}) to evaluate the accuracy of recommendations.
Meanwhile, we use some additional indicators to evaluate the model's ability to follow user instructions:
1). For category control, we use the Top-k target category proportion (\textbf{TCP@K}) metric to evaluate the proportion of target categories in the recommendation list.
The calculation of TCP@K is shown in \autoref{eq:TCP}.
2). For category proportion control, we use the target category proportion accuracy (\textbf{CPA}), defined in \autoref{eq:CPA}, to measure whether the responding item distribution complies with the instruction. Note that $k$ indicates the number of recommended items and $m$ indicates the proportion requirement in instructions.
\begin{equation}
\label{eq:TCP}
TCP@K = \frac{1}{K}\sum_{i=1}^{K}{\mathds{1}(item_i \in C_{target})}
\end{equation}

\begin{equation}
\label{eq:CPA}
\resizebox{0.43\textwidth}{!}{
$CPA =
\begin{cases}
\mathds{1}(Count_{in} \leq k*m),  & \text{if $I_2^{CP\leq m}$} \\
\mathds{1}(Count_{in} \geq k*m),  & \text{if $I_2^{CP\geq m}$} \\
\mathds{1}(\mathrm{abs}(Count_{in}-k*m) \leq 1),  & \text{if $I_2^{CP\approx m}$} \\
\end{cases}$
}
\end{equation}

\subsubsection{Baselines}
We divide the compared methods into 3 classes: general LLM, fine-tuned LLM, and our method and its variants. For sequential recommendation, we additionally compared with \textbf{SASRec}~\cite{kang2018self} that we used as the teacher model.

\paragraph{General LLM:} We use a closed-source LLM \textbf{GPT-3.5-turbo-0613}~\footnote{\url{https://platform.openai.com/docs/models/gpt-3-5-turbo}} (GPT-3.5 for short) and an open-source LLM \textbf{Llama2-7b-chat}~\cite{touvron2023llamachat}. 
Due to the high cost, all GPT-3.5 results are obtained from the first $1{,}000$ samples of the test set. 

\paragraph{Fine-tuned LLM:}
\textbf{InstructRec}~\cite{zhang2023recommendation} uses $39$ types of recommendation-related instructions constructed from user history and review data. It uses Beam search to obtain the top-k recommendation list during the test phase. Its backbone is 3B Flan-T5-XL.  \textbf{PALR}~\cite{chen2023palr} also uses Llama2-7b-chat as the backbone. We fine-tune both models using our datasets, employing data sample generation techniques as described in their respective papers. For instance, a key distinction between PALR and our supervised instruction tuning task, $I_{0}$, is that PALR does not utilize a teacher recommender model for label augmentation; instead, it relies solely on user behavior history to derive label responses.

\paragraph{Our method and variants:}
\textbf{$\mathbf{Ours_{v1}}$} is a model that only uses sequential recommendation instructions ($I_{0}$) in SL. \textbf{$\mathbf{Ours_{v2}}$} is a model uses all instructions ($I_{\{0,1,2,3,4\}}$) in SL. \textbf{$\mathbf{Ours_{v3}}$} is a model obtained by further fine-tuning $\mathrm{Ours_{v2}}$ with RL, but without the item-level reward. \textbf{$\mathbf{Ours_{full}}$} is the complete version of our method. We compare different variants of our method as an ablation study to verify the effect of different components.

\subsection{Overall Performance}

\subsubsection{Sequential Recommendation}
We begin by assessing model performance under standard sequential recommendation scenarios, where users have no explicit intentions. It is important to clarify that our objective in this paper is to improve the ability of LLMs to follow recommendation-related instructions, not to outperform the teacher model, SASRec. Achieving recommendation accuracy on par with the teacher model is considered satisfactory for our purposes.  
\autoref{tab:SeqRecResult} presents the comprehensive results. Initially, all fine-tuned models surpass the performance of general LLMs such as GPT-3.5 and Llama2-7b, affirming the necessity of fine-tuning for domain-specific tasks. Additionally, our approach significantly outperforms the fine-tuned benchmarks (InstructRec and PALR), validating the efficacy of our Supervised Learning (SL) stage. Lastly, our completed model, $\mathbf{Ours_{full}}$, achieves results comparable to the teacher model, SASRec, reinforcing the premise that an LLM must first grasp user preferences before its instruction-following capabilities can be further evaluated.
\begin{table}[t]
\setlength{\abovecaptionskip}{0.1cm}
\resizebox{0.5\textwidth}{!}{
\begin{tabular}{llcccc}
\toprule
Dataset & Method & $\mathrm{HR@10}$ & $\mathrm{NDCG@10}$ & $\mathrm{HR@5}$ & $\mathrm{NDCG@5}$ \\
\midrule
\multirow{8}{*}{Movie}
& $\mathrm{SASRec}$ & $\textbf{0.1229}$ & $\underline{0.0913}$ & $\underline{0.1018}$ & $\underline{0.0844}$ \\
& $\mathrm{GPT-3.5}$ & $0.0050$ & $0.0025$ & $0.0030$ & $0.0019$ \\
& $\mathrm{Llama2-7b}$ & $0.0120$ & $0.0056$ & $0.0133$ & $0.0072$ \\
& $\mathrm{InstructRec}$ & $0.0524$ & $0.0381$ & $0.0406$ & $0.0343$ \\
& $\mathrm{PALR}$ & $0.0868$ & $0.0787$ & $0.0832$ & $0.0775$ \\
\cline{2-6}
& $\mathrm{Ours_{v1}}$ & $0.1108$ & $0.0861$ & $0.0991$ & $0.0827$ \\
& $\mathrm{Ours_{v2}}$ & $\underline{0.1211}$ & $\textbf{0.0927}$ & $\textbf{0.1056}$ & $\textbf{0.0880}$ \\
& $\mathrm{Ours_{v3}}$ & $0.1150$ & $0.0858$ & $0.1010$ & $0.0824$ \\
& $\mathrm{Ours_{full}}$ & $0.1148$ & $0.0867$ & $0.1001$ & $0.0825$ \\
\hline
\hline
\multirow{8}{*}{Steam}
& $\mathrm{SASRec}$ & $\textbf{0.1121}$ & $\textbf{0.0648}$ & $\textbf{0.0778}$ & $\textbf{0.0538}$ \\
& $\mathrm{GPT-3.5}$ & $0.0160$ & $0.0079$ & $0.0090$ & $0.0055$ \\
& $\mathrm{Llama2-7b}$ & $0.0052$ & $0.0028$ & $0.0016$ & $0.0012$ \\
& $\mathrm{InstructRec}$ & $0.0220$ & $0.0113$ & $0.0122$ & $0.0082$ \\
& $\mathrm{PALR}$ & $0.0408$ & $0.0320$ & $0.0373$ & $0.0308$ \\
\cline{2-6}
& $\mathrm{Ours_{v1}}$ & $0.0930$ & $0.0535$ & $0.0666$ & $0.0451$ \\
& $\mathrm{Ours_{v2}}$ & $\underline{0.1036}$ & $\underline{0.0583}$ & $\underline{0.0717}$ & $\underline{0.0479}$ \\
& $\mathrm{Ours_{v3}}$ & $0.1014$ & $0.0557$ & $0.0695$ & $0.0454$ \\
& $\mathrm{Ours_{full}}$ & $0.1001$ & $0.0551$ & $0.0688$ & $0.0449$ \\
\bottomrule
\end{tabular}
}
\caption{Results of ($I_0$). The best result is highlighted in \textbf{boldface} and the runner-up is denoted with \underline{underline}.}
\label{tab:SeqRecResult}
\vspace{-0.3cm}
\end{table}

\subsubsection{Category Control}
\label{sssec:4_2_2}
To evaluate the Category Control Instructions ($\bm{I_1}$), we use the following two settings:

\textbf{Positive ($I_1^{+C}$)}: 
We utilize the target item's category as the control signal $C_{target}$ and incorporate a positive descriptor of $C_{target}$ into the input instruction to represent the user's explicit intent to receive more items within that category.

 \textbf{Negative ($I_1^{-C}$)}: 
We analyze the output statistics of top-$k$ recommendations by SASRec to identify the category with the largest proportion — excluding the target item's category — as $C_{target}$ for simulated control. A negative descriptor of $C_{target}$ is then embedded in the instruction to reflect the user's intent to minimize items from this category.

\newcommand{\contrastive}{\hspace*{.4in}\rotatebox[origin=c]{180}{$\Lsh$}\xspace}

\begin{table*}[t]
\setlength{\abovecaptionskip}{0.1cm}
\centering
\resizebox{14cm}{!}{
\begin{tabular}{ll|ccc|ccc}
\toprule
\multicolumn{2}{c}{$\bm{Control}$} & \multicolumn{3}{c}{$\bm{I_1^{+C}}$} & \multicolumn{3}{c}{$\bm{I_1^{-C}}$} \\
\cmidrule(lr){1-2}\cmidrule(lr){3-5}\cmidrule(lr){6-8}
Dataset & Model & $\mathrm{HR@10}$ & $\mathrm{NDCG@10}$ & $\mathrm{TCP@10(\%)} \uparrow$ & $\mathrm{HR@10}$ & $\mathrm{NDCG@10}$ & $\mathrm{TCP@10(\%)} \downarrow$ \\
\hline
\multirow{7}{*}{\rotatebox[origin=c]{0}{Movie}} 
& $\mathrm{GPT-3.5}$ & $0.0200$ & $0.0111$ & $7.39 (2.23)$ & $0.0150$ & $0.0074$ & $\textbf{7.78} (19.72)$ \\
& $\mathrm{Llama2-7b}$ & $0.0238$ & $0.0109$ & $4.89 (2.80)$ & $0.0259$ & $0.0123$ & $11.61 (18.52)$ \\
& $\mathrm{InstructRec}$ & $0.1045$ & $0.0687$ & $37.39 (7.16)$ & $0.0362$ & $0.0242$ & $56.61 (41.82)$ \\
& $\mathrm{PALR}$ & $0.0832$ & $0.0746$ & $8.70 (7.64)$ & $0.0809$ & $0.0728$ & $46.64 (61.46)$ \\
\cline { 2 - 8 }
& $\mathrm{Ours_{v1}}$ & $0.1054$ & $0.0788$ & $8.69 (8.07)$ & $\underline{0.1023}$ & $\underline{0.0778}$ & $35.99 (40.83)$ \\
& $\mathrm{Ours_{v2}}$ & $\textbf{0.2620}$ & $\textbf{0.1814}$ & $81.38 (8.05)$ & $\textbf{0.1099}$ & $\textbf{0.0832}$ & $19.35(39.80)$ \\
& $\mathrm{Ours_{v3}}$ & $\underline{0.2383}$ & $\underline{0.1609}$ & $\underline{89.06} (7.89)$ & $0.1017$ & $0.0772$ & ${15.80} (39.04)$ \\
& $\mathrm{Ours_{full}}$ & $0.2336$ & $0.1574$ & $\textbf{93.52} (7.81)$ & $0.0999$ & $0.0756$ & $\underline{11.49} (38.41)$ \\
\hline
\hline
\multirow{7}{*}{\rotatebox[origin=c]{0}{Steam}} 
& $\mathrm{GPT-3.5}$ & $0.0360$ & $0.0187$ & $37.62 (25.59)$ & $0.0090$ & $0.0044$ & $19.67 (43.59)$ \\
& $\mathrm{Llama2-7b}$ & $0.0073$ & $0.0044$ & $21.68 (19.94)$ & $0.0047$ & $0.0024$ & $32.57 (50.69)$ \\
& $\mathrm{InstructRec}$ & $0.0494$ & $0.0247$ & $51.70 (29.47)$ & $0.0114$ & $0.0059$ & $70.39 (53.49)$ \\
& $\mathrm{PALR}$ & $0.0392$ & $0.0301$ & $13.04 (12.45)$ & $0.0350$ & $0.0275$ & $47.01 (52.60)$ \\
\cline { 2 - 8 }
& $\mathrm{Ours_{v1}}$ & $0.0922$ & $0.0509$ & $27.23 (26.41)$ & $0.0887$ & $0.0488$ & $51.38 (56.63)$ \\
& $\mathrm{Ours_{v2}}$ & $\textbf{0.3484}$ & $\textbf{0.2110}$ & $92.25 (28.02)$ & $\textbf{0.1236}$ & $\textbf{0.0724}$ & $12.18 (55.67)$ \\
& $\mathrm{Ours_{v3}}$ & $\underline{0.3422}$ & $\underline{0.2049}$ & $\underline{95.13} (29.75)$ & $\underline{0.1200}$ & $\underline{0.0696}$ & $\underline{6.42} (54.37)$ \\
& $\mathrm{Ours_{full}}$ & $0.3395$ & $0.2036$ & $\textbf{95.80} (29.92)$ & $0.1167$ & $0.0676$ & $\textbf{5.51} (54.00)$ \\
\hline
\end{tabular}
}
\caption{Results of $I_1$ control. The best result is highlighted in \textbf{boldface} and the runner-up is denoted with \underline{underline}. Values in parentheses show outcomes of the corresponding model without control signals.}
\label{tab:PersonalControlRecResult}
\end{table*} 
\autoref{tab:PersonalControlRecResult} presents the results, using the TCP@10 metric to assess conformity to control signals, while HR@10 and NDCG@10 indicate recommendation accuracy when explicit intentions are additionally included in the prompt. Values in parentheses show the outcomes of each corresponding model without control signals. Two key observations emerge: Firstly, differentiating supervised learning tasks by intention is essential. Evidence for this includes InstructRec's improved TCP performance in the $I_1{+C}$ task, which aligns with its training, versus no effect in the $I_1{-C}$ task not covered by its training. Similarly, $\mathbf{Ours_{v1}}$, trained solely on recommendation tasks, underperforms in TCP compared to $\mathbf{Ours_{v2}}$. Secondly, $\mathbf{Ours_{v2}}$, $\mathbf{Ours_{v3}}$, and $\mathbf{Ours_{full}}$ show notable TCP enhancements, with $\mathbf{Ours_{full}}$ leading. Additionally, HR and NDCG metrics also see a significant lift compared to \autoref{tab:SeqRecResult}, confirming our method's effectiveness in adhering to category control instructions.

\subsubsection{Category Proportion Control} 
To evaluate the Category Control Instructions ($\bm{I_1}$), we use the following three settings:

\textbf{$\bm{I_2^{CP\leq20\%}}$}:  
In this case, we select the $C_{target}$ of $I_1^{-C}$ in the same way to simulate control. The purpose is to control the proportion of $C_{target}$ to be no greater than 20\%.

\textbf{$\bm{I_2^{CP\approx30\%}}$}:  
In this case, we select the category of the target item as the $C_{target}$ of control simulating. The purpose is to control the proportion of items in $C_{target}$ to be approximately 30\%.

\textbf{$\bm{I_2^{CP\geq30\%}}$}:  
We select the category of the target item as the $C_{target}$. The purpose is to control the proportion of items in $C_{target}$ to be no less than 30\%.
 
Results are reported in \autoref{tab:PersonalCategoryRateResult}. In this case, the primary metric is CPA. Overall, Ours$_{full}$ achieves the best performance, outperforming all other baselines and its own variants.

\begin{table*}[t]
\setlength{\abovecaptionskip}{0.1cm}
\resizebox{16cm}{!}{
\begin{tabular}{ll|ccc|ccc|ccc}
\toprule
\multicolumn{2}{c}{$\bm{Control}$} & \multicolumn{3}{c}{$\bm{I_2^{CP\leq20\%}}$} & \multicolumn{3}{c}{$\bm{I_2^{CP\approx30\%}}$} & \multicolumn{3}{c}{$\bm{I_2^{CP\geq30\%}}$} \\
\cmidrule(lr){1-2}\cmidrule(lr){3-5}\cmidrule(lr){6-8}\cmidrule(lr){9-11}
Dataset & Model & $\mathrm{HR@10}$ & $\mathrm{NDCG@10}$ & $\mathrm{CPA(\%)}\uparrow$ & $\mathrm{HR@10}$ & $\mathrm{NDCG@10}$ & $\mathrm{CPA(\%)} \uparrow$ & $\mathrm{HR@10}$ & $\mathrm{NDCG@10}$ & $\mathrm{CPA(\%)}\uparrow$ \\
\hline
\multirow{7}{*}{\rotatebox[origin=c]{0}{Movie}} 
& $\mathrm{GPT-3.5}$ & $0.0050$ & $0.0021$ & $1.80(0.50)$ & $0.0210$ & $0.0124$ & $9.71(0.70)$ & $0.0250$ & $0.0167$ & $6.61(1.10)$ \\
& $\mathrm{Llama2-7b}$ & $0.0198$ & $0.0089$ & $16.30(9.83)$ & $0.0270$ & $0.0114$ & $9.71(1.14)$ & $0.0279$ & $0.0119$ & $3.45(2.09)$ \\
& $\mathrm{InstructRec}$ & $0.0372$ & $0.0252$ & $9.75(15.87)$ & $0.1031$ & $0.0684$ & $31.24(3.96)$ & $0.1041$ & $0.0692$ & $\underline{57.58}(9.60)$ \\
& $\mathrm{PALR}$ & $0.0792$ & $0.0711$ & $29.76(6.70)$ & $0.0813$ & $0.0724$ & $10.28(3.16)$ & $0.0825$ & $0.0733$ & $11.17(9.84)$ \\
\cline { 2 - 11 }
& $\mathrm{Ours_{v1}}$ & $\underline{0.1018}$ & $\textbf{0.0742}$ & $30.97(14.34)$ & $0.1048$ & $0.0766$ & $16.14(4.43)$ & $0.1061$ & $0.0776$ & $10.46(10.19)$ \\
& $\mathrm{Ours_{v2}}$ & $\textbf{0.1027}$ & $\underline{0.0688}$ & $29.18(14.93)$ & $\textbf{0.2034}$ & $\textbf{0.1417}$ & $58.71(4.61)$ & $\underline{0.2055}$ & $\underline{0.1433}$ & $45.89(9.96)$ \\
& $\mathrm{Ours_{v3}}$ & $0.0943$ & $0.0594$ & $\underline{33.74}(15.03)$ & $0.1948$ & $0.1319$ & $\underline{65.78}(4.58)$ & $0.1994$ & $0.1340$ & $48.19(9.21)$ \\
& $\mathrm{Ours_{full}}$ & $0.0981$ & $0.0642$ & $\textbf{45.90}(16.33)$ & $\underline{0.2005}$ & $\underline{0.1354}$ & $\textbf{69.81}(4.68)$ & $\textbf{0.2134}$ & $\textbf{0.1447}$ & $\textbf{61.98}(9.15)$ \\
\hline
\hline
\multirow{7}{*}{\rotatebox[origin=c]{0}{Steam}} 
& $\mathrm{GPT-3.5}$ & $0.0170$ & $0.0074$ & $29.80(14.40)$ & $0.0290$ & $0.0152$ & $25.60(3.40)$ & $0.0360$ & $0.0202$ & $51.10(39.60)$ \\
& $\mathrm{Llama2-7b}$ & $0.0041$ & $0.0022$ & $14.65(7.24)$ & $0.0024$ & $0.0011$ & $23.06(6.79)$ & $0.0026$ & $0.0013$ & $32.97(34.65)$ \\
& $\mathrm{InstructRec}$ & $0.0109$ & $0.0056$ & $15.86(21.39)$ & $0.0442$ & $0.0223$ & $13.22(3.15)$ & $0.0435$ & $0.0220$ & $55.90(40.30)$ \\
& $\mathrm{PALR}$ & $0.0333$ & $0.0257$ & $6.18(2.47)$ & $0.0374$ & $0.0287$ & $13.39(4.13)$ & $0.0388$ & $0.0296$ & $20.11(18.96)$ \\
\cline { 2 - 11 }
& $\mathrm{Ours_{v1}}$ & $0.0863$ & $0.0478$ & $20.51(12.25)$ & $0.0897$ & $0.0496$ & $12.23(3.96)$ & $0.0903$ & $0.0502$ & $39.22(38.73)$ \\
& $\mathrm{Ours_{v2}}$ & $0.1144$ & $0.0647$ & $41.26(14.90)$ & $0.2072$ & $0.1191$ & $71.97(3.26)$ & $0.2184$ & $0.1238$ & $70.79(39.79)$ \\
& $\mathrm{Ours_{v3}}$ & $\underline{0.1172}$ & $\underline{0.0663}$ & $\underline{65.41}(18.04)$ & $\underline{0.2105}$ & $\underline{0.1206}$ & $\textbf{77.51}(2.14)$ & $\underline{0.2283}$ & $\underline{0.1286}$ & $\underline{82.07}(40.69)$ \\
& $\mathrm{Ours_{full}}$ & $\textbf{0.1183}$ & $\textbf{0.0663}$ & $\textbf{70.58}(18.15)$ & $\textbf{0.2225}$ & $\textbf{0.1295}$ & $\underline{74.29}(2.77)$ & $\textbf{0.2382}$ & $\textbf{0.1365}$ & $\textbf{88.40}(40.98)$ \\
\hline
\end{tabular}
}
\caption{Results of $I_2$ control. The best result is highlighted in \textbf{boldface} and the runner-up is denoted with \underline{underline}. Values in parentheses show outcomes of the corresponding model without control signals.}
\label{tab:PersonalCategoryRateResult}
\vspace{-0.3cm}
\end{table*}

\subsection{Formatting and General Evaluation} 
Finally, we assess the formatting ability and overall linguistic capabilities of LLMs in generating structured recommendations. We measure formatting ability for the recommendation domain in four dimensions: 1) \textbf{CorrectCount}: The accuracy of the recommended item count against the given $k$ in the instruction. 2) \textbf{RepeatItem}: The frequency of repeated items in the recommendation list. 3) \textbf{NonExist}: The occurrence of non-existent (hallucinated) items in the list. 4) \textbf{InHistory}: The rate of recommended items already present in the user's history. For a more challenging test, LLMs are tasked with recommending $k$ items where $k$ is a random number between 11 and 15, despite training only on ranges from 1 to 10.
To evaluate the general language ability of LLMs, we utilize two standard tasks, \textbf{MMLU} with five examples (5-shot) and \textbf{GSM8K} with eight examples (8-shot), leveraging the unified framework provided by \url{https://github.com/EleutherAI/lm-evaluation-harness}.
 

\begin{table*}[t]
\setlength{\abovecaptionskip}{0.1cm}
\resizebox{16cm}{!}{
\begin{tabular}{l|l|cccc|cc|cc}
\toprule
\multirow{2}{*}{Dataset} & \multirow{2}{*}{Method} & \multicolumn{4}{c}{Formatting Quality(\%)} & \multicolumn{2}{|c}{Precision} & \multicolumn{2}{|c}{Generalization} \\
\cmidrule(lr){3-6}\cmidrule(lr){7-8}\cmidrule(lr){9-10}
& & $\mathrm{CorrectCount}\uparrow$ & $\mathrm{RepeatItem@K}\downarrow$ & $\mathrm{NonExist@K}\downarrow$ & $\mathrm{InHistory@K}\downarrow$ & $\mathrm{HR@K}\uparrow$ & $\mathrm{NDCG@K}\uparrow$ & $\mathrm{MMLU}\uparrow$ & $\mathrm{GSM8K}\uparrow$\\
\midrule
\multirow{8}{*}{Movie}
& $\mathrm{GPT-3.5}$ & $100.00$ & $2.45$ & $66.10$ & $4.22$ & $0.0100$ & $0.0034$ & $0.700$ & $0.7460$ \\
& $\mathrm{Llama2-7b}$ & $99.75$ & $5.64$ & $49.47$ & $29.74$ & $0.0183$ & $0.0075$ & $0.440$ & $\textbf{0.2858}$ \\
& $\mathrm{InstructRec}$ & $100.00$ & $10.01$ & $8.52$ & $15.35$ & $0.0546$ & $0.0378$ & $-$ & $-$ \\
& $\mathrm{PALR}$ & $77.27$ & $65.92$ & $4.06$ & $9.31$ & $0.0869$ & $0.0787$ & $0.377$ & $0.1099$ \\
\cline { 2 - 10 }
& $\mathrm{Ours_{v1}}$ & $92.96$ & $13.63$ & $7.03$ & $5.76$ & $0.1164$ & $0.0876$ & $0.341$ & $0.1318$ \\
& $\mathrm{Ours_{v2}}$ & $100.00$ & $9.61$ & $5.27$ & $4.02$ & $\textbf{0.1285}$ & $\textbf{0.0950}$ & $0.450$ & $\underline{0.1842}$ \\
& $\mathrm{Ours_{v3}}$ & $\underline{100.00}$ & $\underline{2.37}$ & $\underline{1.14}$ & $\underline{1.36}$ & $0.1214$ & $0.0886$ & $\underline{0.453}$ & $0.1789$ \\
& $\mathrm{Ours_{full}}$ & $\textbf{100.00}$ & $\textbf{1.14}$ & $\textbf{0.95}$ & $\textbf{1.24}$ & $\underline{0.1220}$ & $\underline{0.0890}$ & $\textbf{0.455}$ & $0.1782$ \\
\midrule
\midrule
\multirow{8}{*}{Steam}
& $\mathrm{GPT-3.5}$ & $99.90$ & $2.44$ & $26.76$ & $4.50$ & $0.0200$ & $0.0094$ & $0.700$ & $0.7460$ \\
& $\mathrm{Llama2-7b}$ & $99.79$ & $5.88$ & $20.78$ & $43.90$ & $0.0074$ & $0.0030$ & $0.440$ & $\textbf{0.2858}$ \\
& $\mathrm{InstructRec}$ & $98.41$ & $0.99$ & $4.60$ & $7.54$ & $0.0270$ & $0.0130$ & $-$ & $-$ \\
& $\mathrm{PALR}$ & $97.53$ & $17.67$ & $46.78$ & $2.88$ & $0.0404$ & $0.0316$ & $0.417$ & $0.1327$ \\
\cline { 2 - 10 }
& $\mathrm{Ours_{v1}}$ & $95.79$ & $3.95$ & $3.00$ & $1.49$ & $0.1029$ & $0.0559$ & $0.327$ & $0.0819$ \\
& $\mathrm{Ours_{v2}}$ & $100.00$ & $2.68$ & $1.59$ & $2.55$ & $\textbf{0.1152}$ & $\textbf{0.0612}$ & $\textbf{0.458}$ & $\underline{0.2146}$ \\
& $\mathrm{Ours_{v3}}$ & $\underline{100.00}$ & $\underline{0.37}$ & $\underline{1.04}$ & $\underline{0.22}$ & $0.1149$ & $\underline{0.0593}$ & $0.457$ & $0.2039$ \\
& $\mathrm{Ours_{full}}$ & $\textbf{100.00}$ & $\textbf{0.23}$ & $\textbf{0.78}$ & $\textbf{0.17}$ & $\underline{0.1149}$ & $0.0589$ & $\underline{0.457}$ & $0.2123$ \\
\bottomrule
\end{tabular}
}
\caption{Results of formatting and general evaluation. The best result (excluding GPT-3.5) is highlighted in \textbf{boldface} and the runner-up is denoted with \underline{underline}.}
\label{tab:AblationResult}
\end{table*}
\autoref{tab:AblationResult} presents the comprehensive outcomes. Notably, nearly all models, with the exception of PALR on the Movie dataset, excel in the CorrectCount metric. For other formatting metrics such as RepeatItem, NonExist, and InHistory, a progressive enhancement is evident from Ours$_{v1}$ through Ours$_{v3}$, culminating in Ours$_{full}$. In \autoref{tab:AblationResult}, the primary focus is on \textsl{Formatting Quality}, while \textsl{Precision} metrics serve as supplementary references. The \textsl{Generalization} metrics indicate the extent of catastrophic forgetting. When compared to the fine-tuned baseline PALR, Ours$_{full}$ shows the least performance deterioration from its underlying Llama2-7b architecture. Owing to the 512-token input limitation of InstructRec, its performance could not be assessed on the \textsl{Generalization} tasks.

\subsection{Case Study}\label{sec:case}

\begin{table*}[t]
\setlength{\abovecaptionskip}{0.1cm}
\resizebox{16cm}{!}{
\begin{tabular}{l|l}
\toprule
\multicolumn{2}{p{27cm}}{\textbf{Instruction $\mathbf{I_2^{CP\approx50\%}}$: } You are an expert recommender engine. You need to generate a recommendation list simultaneously considering user‘s preference and user’s intention. Here are the user's historical interactions: \textcolor{ForestGreen}{'The Blind Swordsman: Zatoichi Takeshi Kitano; Tadanobu Asano japan', 'Buffy the Vampire Slayer - The Complete Seventh Season', 'Number Ones', '24: Season 3', 'Madonna - The Immaculate Collection', 'Spider-Man 2', 'Heavy Metal 2000 VHS', 'MI-5, Vol. 1', 'MI-5, Vol. 2', 'Beyond Re Animator VHS'}, and user's intention: \textcolor{BurntOrange}{In the recommendation list, the proportion of \textbf{'Art House \& International, By Original Language, Chinese'} items should be approximately 50\%}. You need to generate a recommendation list with \underline{5} different items.} \\
\midrule
$\mathbf{GPT-3.5}$ & 1. Oldboy 2. Chungking Express 3. Battle Royale 4. In the Mood for Love 5. Memories of Murder \\
\midrule
$\mathbf{Llama2-7b}$ & 1. \colorbox{lightgray}{The Blind Swordsman: Zatoichi} 2. \colorbox{lightgray}{Buffy the Vampire Slayer} 3. \colorbox{lightgray}{Number Ones} 4. \colorbox{lightgray}{24: Season 3} 5. \colorbox{lightgray}{Madonna - The Immaculate Collection} \\
\midrule
$\mathbf{InstructRec}$ & 1. The Magnificent Seven VHS 2. Shanghai Noon 3. X-Men: First Class 4. X-Men VHS 5. Akira VHS \\
\midrule
$\mathbf{Ours_{full}}$ & 1. \colorbox{YellowGreen}{\textbf{\underline{Five Deadly Venoms VHS}}} 2. \colorbox{YellowGreen}{Ip Man 2} 3. The Village: Vista Series 4. Dawn of the Dead 5. \colorbox{YellowGreen}{Fong Sai Yuk VHS} \\
\midrule
\midrule
\multicolumn{2}{p{27cm}}{\textbf{Instruction $\mathbf{I_1^{-C}}$: } You are an expert recommender engine. You need to generate a recommendation list simultaneously considering user‘s preference and user’s intention. Here are the user's historical interactions: \textcolor{ForestGreen}{'Yet Another Zombie Defense HD', 'Battlefield™ V', 'Tom Clancy's Ghost Recon® Wildlands', 'Agents of Mayhem', 'Beholder', 'HELLDIVERS™ Dive Harder Edition', 'SUPERHOT', 'Watch\_Dogs® 2', 'Deus Ex: Mankind Divided', 'POSTAL 2'}, and user's intention: \textcolor{BurntOrange}{Don't recommend me any \textbf{'Shooter'} products}. You need to generate a recommendation list with \underline{5} different items.} \\
\midrule
$\mathbf{GPT-3.5}$ & 1. Stardew Valley 2. Hollow Knight 3. Cities: Skylines 4. Divinity: Original Sin 2 5. The Witness \\
\midrule
$\mathbf{Llama2-7b}$ & 1. Assassin's Creed Odyssey 2. The Witcher 3: Wild Hunt 3. Horizon Zero Dawn 4. \colorbox{YellowGreen}{Far Cry 5} 5. Life is Strange 2 \\
\midrule
$\mathbf{InstructRec}$ & 1. \colorbox{lightgray}{POSTAL 2} 2. \colorbox{YellowGreen}{POSTAL Redux} 3. \colorbox{YellowGreen}{Half-Life 2} 4. \colorbox{YellowGreen}{Resident Evil 0} 5. \colorbox{YellowGreen}{POSTAL} \\
\midrule
$\mathbf{Ours_{full}}$ & 1. The Witcher® 3: Wild Hunt 2. Human: Fall Flat 3. \textbf{\underline{The Forest}} 4. Outlast 5. Middle-earth™: Shadow of War™  \\
\bottomrule
\end{tabular}
}
\caption{Two real examples for case study. We highlight some texts for \colorbox{lightgray}{Repeat with history item}, \colorbox{YellowGreen}{In target category item} and \textbf{\underline{target item}}.}
\label{tab:CaseStudy}
\end{table*}
For clarity, we present two illustrative cases in \autoref{tab:CaseStudy}. The first is about the $I_2^{CP\approx50\%}$ instruction on the Movie dataset. A user requests that roughly 50\% of the top-5 recommendations feature ``Art House \& International, By Original Language, Chinese'' characteristics. Our method successfully generates a list where 3 out of 5 items possess these attributes, while three baseline methods fail to meet this criterion. Llama-2-7b simply repeats items mentioned in the user history.

In the second case, concerning the $I_1^{-C}$ instruction, the user specifies to exclude ``Shooter'' games from recommendations. Our method effectively adheres to this restriction by omitting any items from the blacklisted category and successfully includes the ground-truth item in the recommendation list. Conversely, Llama-2-7b incorrectly suggests ``Far Cry 5'', a Shooter game, while InstructRec not only makes aching historical mistakes but also recommends a lot of games from the shooter genre.


\begin{table*}[t]  
\setlength{\abovecaptionskip}{0.1cm}  
\centering  
\resizebox{12cm}{!}{  
\begin{tabular}{ll|cc|ccc}  
\toprule  
\multicolumn{2}{c}{$\bm{Control}$} & \multicolumn{2}{c}{$I_1^{+C_1} \ \& \   I_1^{-C_2}$} & \multicolumn{3}{c}{$I_2^{C_1P\leq 20\%}  \ \& \  I_1^{-C_2}$} \\  
\cmidrule(lr){1-2}\cmidrule(lr){3-4}\cmidrule(lr){5-7}  
Dataset & Model & $\mathrm{HR@10}$ & $\mathrm{TC_{1 \neg 2}P@10(\%)} \uparrow$ & $\mathrm{HR@10}$ & $C_1PA \uparrow$ & $\mathrm{TC_2P@10(\%)} \downarrow$ \\  
\hline  
\multirow{8}{*}{\rotatebox[origin=c]{0}{Movie}}   
& $\mathrm{GPT-3.5}$ & $0.0176$ & $7.15 (2.20)$ & $0.0132$ & $9.84 (6.37)$ & $\textbf{12.51} (19.45)$ \\  
& $\mathrm{Llama2}$ & $0.0155$ & $4.51 (2.79)$ & $0.0226$ & $16.50 (10.34)$ & $17.04 (23.82)$ \\  
& $\mathrm{InstructRec}$ & $0.0944$ & $27.23 (7.14)$ & $0.0333$ & $12.83 (9.47)$ & $51.81 (31.91)$ \\  
& $\mathrm{PALR}$ & $0.0762$ & $7.99 (7.60)$ & $0.0692$ & $33.21 (6.79)$ & $38.08 (53.46)$ \\  
\cline { 2 - 7 }  
& $\mathrm{Ours_{v1}}$ & $0.1002$ & $8.21 (8.05)$ & $0.0965$ & $28.40 (11.94)$ & $28.69 (31.27)$ \\  
& $\mathrm{Ours_{v2}}$ & $0.2031$ & $34.63 (8.04)$ & $\underline{0.1012}$ & $29.13 (13.21)$ & $25.28 (30.40)$ \\  
& $\mathrm{Ours_{v3}}$ & $\underline{0.2033}$ & $\underline{44.16} (7.88)$ & $0.0927$ & $\underline{36.25} (15.07)$ & $22.01 (28.88)$ \\  
& $\mathrm{Ours_{full}}$ & $\textbf{0.2064}$ & $\textbf{49.61} (7.89)$ & $\textbf{0.1018}$ & $\textbf{45.14} (15.95)$ & $\underline{16.65} (28.50)$ \\  
\hline  
\multirow{8}{*}{\rotatebox[origin=c]{0}{Steam}}   
& $\mathrm{GPT-3.5}$ & $0.0194$ & $34.15 (21.85)$ & $0.0183$ & $15.38 (4.21)$ & $\underline{22.13} (34.18)$ \\  
& $\mathrm{Llama2}$ & $0.0093$ & $18.14 (16.04)$ & $0.0073$ & $12.49 (4.99)$ & $30.40 (38.49)$ \\  
& $\mathrm{InstructRec}$ & $0.0432$ & $27.94 (18.63)$ & $0.0133$ & $13.84 (20.37)$ & $38.23 (34.63)$ \\  
& $\mathrm{PALR}$ & $0.0353$ & $10.52 (11.09)$ & $0.0294$ & $7.04 (2.53)$ & $41.31 (47.31)$ \\  
\cline { 2 - 7 }  
& $\mathrm{Ours_{v1}}$ & $0.0859$ & $25.07 (25.31)$ & $0.0842$ & $21.58 (12.25)$ & $36.18 (41.81)$ \\  
& $\mathrm{Ours_{v2}}$ & $0.2620$ & $42.39 (26.95)$ & $\underline{0.1187}$ & $54.06 (14.58)$ & $29.63 (39.34)$ \\  
& $\mathrm{Ours_{v3}}$ & $\underline{0.2729}$ & $\underline{48.27} (28.83)$ & $0.1183$ & $\underline{75.56} (17.74)$ & $22.38 (35.74)$ \\  
& $\mathrm{Ours_{full}}$ & $\textbf{0.2807}$ & $\textbf{56.00} (28.99)$ & $\textbf{0.1204}$ & $\textbf{77.26} (17.81)$ & $\textbf{18.82} (35.56)$ \\  
\hline  
\end{tabular}  
}  
\caption{Results of two combinatorial control instructions: (1) $I_1^{+C_1} \ \& \   I_1^{-C_2}$ ; (2) $ I_2^{C_1P\leq20\%}  \ \& \  I_1^{-C_2} $. The best result is highlighted in \textbf{boldface} and the runner-up is denoted with \underline{underline}. Values in parentheses show outcomes of the corresponding model without control signals.}  
\label{tab:MultiControl}  
\end{table*}  

\subsection{Extend to combinatorial Controls}\label{sec:multi_control}
Considering LLMs' strong generalization capabilities, we further examine the performance of LLMs on complex, combinatorial instructions not encountered during training, as shown in \autoref{tab:MultiControl}. Here, $TC_{1 \neg 2}P@10(\%)$ means the percentage of items belong to $C_1$ but does not belong to $C_2$. The results confirm our expectations that while our methods surpass baseline models, they achieve lower compliance with user instructions, as measured by TCP and CPA, compared to simpler single-control instructions. This motivates us to include some complex instructions in the alignment process in future work.

\section{Related Work}

In recent years, the remarkable natural language processing capabilities of LLMs have inspired researchers to leverage them for recommendation tasks \cite{fan2023recommender}. Early implementations utilized language models as feature extractors, creating knowledge-aware recommendation embeddings exemplified by models such as U-BERT \cite{qiu2021u} and UserBERT \cite{wu2022userbert}. With the advent of generative models like GPT, the focus has shifted towards generative recommendation models that frame recommendations as natural language generation tasks \cite{wu2023survey}.

Initially, adaptation of LLMs to recommendation scenarios relied heavily on techniques like prompt engineering \cite{gao2023chat,sun2023chatgpt} and contextual learning \cite{dai2023uncovering,liu2023chatgpt}. However, these LLMs often underperform compared to traditional recommendation models trained on task-specific data, prompting the necessity to fine-tune LLMs for better alignment with recommendation tasks. P5 \cite{geng2022recommendation} introduced a unified framework that integrates 5 recommendation tasks through fine-tuning on the FLAN-T5 model \cite{raffel2020exploring}. Subsequently, InstructRec \cite{zhang2023recommendation} tailored FLAN-T5 for various downstream recommendation tasks using instruction tuning. TALLRec \cite{bao2023tallrec} fine-tunes LLaMA for recommendations with very few training samples, but it focuses on the binary classification task. PALR \cite{chen2023palr} employs two types of instructions for instruction tuning to facilitate list-level recommendation generation.

Despite these advancements, current studies have not fully explored the potential of LLMs to enhance the interactivity of recommender systems. We aim to harness the instruction-following prowess of LLMs for recommendation tasks through fine-tuning, to create a conversational, controllable, and interactive recommender agent.
\section{Conclusion}
In conclusion, our paper presents a new approach to tailor LLMs for interactive recommender systems, combining a supervised learning phase with innovative control tasks and a reinforcement learning stage with specialized reward signals. Our method successfully meets the detailed demands of recommendation contexts, enhancing LLMs' performance. Experimental results show our approach exceeds current LLM-based systems in precision, controllability, and presentation, offering a significant step towards refined and reliable recommendation services.

\clearpage

\section{Limitation}
This paper's primary constraints are as follows: (1) The emphasis is placed on improving the LLM's ability to follow recommendation-related instructions. This focus may inadvertently compromise the LLM's broader intellectual capabilities, as revealed in \autoref{tab:AblationResult}. How to further reduce catastrophic forgetting remains a big challenge. (2) In real-world scenarios, users often have diverse control intentions, including intricate blends of various instructions (as discussed in Section~\ref{sec:multi_control}) and new instructions beyond category control. As foundational research, this paper addresses only the most critical elements, specifically category control and formatting control. More diverse and complicated instructions are yet to be explored.

\section{Acknowledgments}
This work is supported by the National Natural Science Foundation of China (Grant Nos. 62276171 and 62072311), Guangdong Basic and Applied Basic Research Foundation (Grant Nos. 2024A1515011938, 2020B1515120028), Guangdong Provincial Key Laboratory of Popular High Performance Computers (Grant No. 2017B030314073), Guangdong Province Engineering Center of China-made High Performance Data Computing System, Shenzhen Fundamental Research-General Project (Grant Nos. 20220811155803001, 20210324094402008). Hao Liao is the corresponding author.

%
\bibliography{references.bib}
\clearpage

\appendix
\section{Appendix}

\subsection{Dataset detail}
\label{app:dataset_detail}

We present detailed information about the instruction set we construct in \autoref{tab:DatasetDetail}, including the quantity of each type of instruction.
\begin{table}[h]
\centering
\resizebox{0.5\textwidth}{!}{
\begin{tabular}{l|c|c}
\toprule
Instruct type & Movie & Steam \\
\midrule
$I_0$ & $13218$ & $12658$ \\
$(I_1^{+C}, I_1^{-C})$ & $(6609,6609)$ & $(6329,6329)$ \\
$(I_2^{CP\leq *}, I_2^{CP\geq *}, I_2^{CP\approx *})$ & $(4406,4406,4406)$ & $(4219,4219,4220)$ \\
$I_3$ & $18744$ & $8572$ \\
$I_4$ & $29199$ & $23273$ \\
\midrule
$ALL$ & $87597$ & $69819$ \\
\bottomrule
\end{tabular}
}
\caption{Constructed instruction set}
\label{tab:DatasetDetail}
\vspace{-0.3cm}
\end{table}

\begin{algorithm}[!h]
\footnotesize
\SetKwData{Left}{left}
\SetKwData{This}{this}
\SetKwData{Up}{up} 
\SetKwFunction{Union}{Union}
\SetKwFunction{FindCompress}{FindCompress} 
\SetKwInOut{Input}{input}
\SetKwInOut{Output}{output}
\SetKwInOut{Output}{output}
\SetKwInOut{Init}{init}
\SetKw{Elif}{elif}
\caption{Calculation of $Scores^{ctl}$}
\label{alg:item_wise_reward}
    \Input{$items=[item_1,...,item_N],k,C_{target}$}
	\Output{Each item's score $Scores^{ctl}$}
    \Init {$Scores^{ctl}=\mathbf{0}^{1\times N}$}
    \Init {$Count_{in}=Count_{out}=0$}
	\BlankLine 
	\For{$i\leftarrow 1$ \KwTo $N$}{
        \If{$item_i$ $\mathrm{is}$ $\mathrm{illegal}$}{
            $Scores^{ctl}_i=-1$ \\
            continue
        }
        \lIf{$item_i \in C_{target}$}{$in=1,out=0$}
        \lElse{$in=0,out=1$}
        $Count_{in}+=in,Count_{out}+=out$
        \BlankLine
        \lIf{$I_0$}{$s=Scores_i$}
        \BlankLine
        \lIf{$I_1^{+C}$}{$s=in$}
        \BlankLine
        \lIf{$I_1^{-C}$}{$s=out$}
        \BlankLine
        \If{$I_2^{CP\leq m}$}{
            \lIf{$Count_{out}>(k-k*m)$}{$s=0.5$}
            \lElseIf{$out$}{$s=1.0$}
            \lElseIf{$Count_{in}<k*m$}{$s=0.5$}
            \lElse{$s=0.0$}
        }
        \BlankLine
        \If{$I_2^{CP\geq m}$}{
            \lIf{$Count_{in}>k$}{$s=0.5$}
            \lElseIf{$in$}{$s=1.0$}
            \lElseIf{$Count_{out}<(k-k*m)$}{$s=0.5$}
            \lElse{$s=0.0$}
        }
        \BlankLine
        \If{$I_2^{CP\approx m}$}{
            \If{$in$}{
                \lIf{$Count_{in}\leq k*m$}{$s=1.0$}
                \lElse{$s=0.0$}
            }
            \lElseIf{$Count_{in}\geq k*m$}{$s=1.0$}
            \lElseIf{$Count_{out}\leq(k-k*m)$}{$s=0.5$}
            \lElse{$s=0.0$}
        }
        $Scores^{ctl}_i = s$
    }
\end{algorithm}

\subsection{Computational Cost}
\label{app:compute_cost}
Finetuning LLMs for recommendations is a little bit more expensive than training traditional recommender models like SASRec, but the resource and cost are not a big barrier even for small organizations or researchers. Here we take the Movie dataset for illustration. The dataset statistics can be found in \autoref{tab:DatasetStatistic} and \autoref{tab:MultiControl}.

For the SL stage, we use 4 A100 GPUs (40GB GPU memory) for training. The batch size on each GPU is 1 and the gradient accumulation step of 16. The maximum sequence length is set to 1024.  Under this setting, each epoch costs about 70 minutes. Usually, models can fully converge in 30 epochs. So, the total training time for SFT is about 35 hours. 

For the RL stage, we use 2 A100 GPUs (40GB GPU memory) for training. The batch size on each GPU is 1 and the gradient accumulation step of 2. For each batch, instruction sampling time + 2 candidate response generation time + model training time counts about 40 seconds, and the maximum training steps are set to 3k.  Thus, the total RL training time is $3000*40/60/60=33$ hours. 

Once the training process is finished, we use vllm\footnote{\url{https://github.com/vllm-project/vllm}} for inference and serving. On a single A100 GPU (40GB memory), when responding to the top-10 recommendation request, our implementation can process 20 requests per second (on average each request involves newly generated 112 tokens).

\definecolor{backcolour}{rgb}{0.96, 0.96, 0.96}
\definecolor{codegreen}{rgb}{0,0.6,0}
\lstdefinestyle{myStyle}{
    backgroundcolor=\color{backcolour},   
    commentstyle=\color{codegreen},
    basicstyle=\ttfamily\small,
    breakatwhitespace=true,         
    breaklines=true,                 
    keepspaces=true,                 
    numbers=none,       
    numbersep=5pt,                  
    showspaces=false,                
    showstringspaces=false,
    showtabs=false,                  
    tabsize=2,
    frame=single,
}
\lstset{style=myStyle}

\subsection{Prompts}
\label{app:prompts}
In this section, we show the prompt for the instruction used in the experiment. The prompt of $I_{0},I_{1},I_{2},I_{3}$ is illustrated in \autoref{lst:I_0} to \autoref{lst:I_3} respectively. The prompt of positive and negative category control intentions is illustrated in \autoref{lst:I_1_p} to \autoref{lst:I_1_n} respectively

\onecolumn
\begin{lstlisting}[caption=Prompts of $I_{0}$, label={lst:I_0}]
Instruction: You are an expert recommender engine. You need to generate a recommendation list considering user's preference from historical interactions. The historical interactions are provided as follows: {history}. You need to generate a recommendation list with {item_count} different items.
Output: {item_list}

Instruction: You are an expert recommender engine. You need to select a recommendation list considering user's preference from historical interactions. The historical interactions are provided as follows: {history}. The candidate items are: {candidate_titles}. You need to select a recommendation list with {item_count} different items from candidate items.
Output: {item_list}
\end{lstlisting}

\begin{lstlisting}[caption=Prompts of $I_{1}$, label={lst:I_1}]
Instruction: You are an expert recommender engine. You need to generate a recommendation list simultaneously considering user's preference inferred from historical interactions and user's intention. If user's preference conflicts with his intention, you should comply with his intention. Here are user's historical interactions: {history}, and user's intention: {synthetic_intention}. You need to generate a recommendation list with {item_count} different items. 
Output: {item_list}

Instruction: You are an expert recommender engine. You need to select a recommendation list from candidate items simultaneously considering user's preference inferred from historical interactions and user's intention. If user's preference conflicts with his intention, you should comply with his intention. Here are user's historical interactions: {history}, and user's intention: {synthetic_intention}. The candidate items are: {candidate_titles}. You need to select a recommendation list with {item_count} different items from candidate items.
Output: {item_list}
\end{lstlisting}

\begin{lstlisting}[caption=Prompts of $I_{2}$, label={lst:I_2}]
Instruction: You are an expert recommender engine. You need to generate a recommendation list simultaneously considering user's preference inferred from historical interactions and user's intention. Here are user's historical interactions: {history}, and user's intention: In the recommendation list, the proportion of '{target_category}' items should be less than or equal to {category_proportion}. You need to generate a recommendation list with {item_count} different items.
Output: {item_list}

Instruction: You are an expert recommender engine. You need to generate a recommendation list simultaneously considering user's preference inferred from historical interactions and user's intention. Here are user's historical interactions: {history}, and user's intention: In the recommendation list, the proportion of '{target_category}' items should be more than or equal to {category_proportion}. You need to generate a recommendation list with {item_count} different items.

Instruction: You are an expert recommender engine. You need to generate a recommendation list simultaneously considering user's preference inferred from historical interactions and user's intention. Here are user's historical interactions: {history}, and user's intention: In the recommendation list, the proportion of '{target_category}' items should be approximately {category_proportion}. You need to generate a recommendation list with {item_count} different items.
Output: {item_list}
\end{lstlisting}

\begin{lstlisting}[caption=Prompts of $I_{3}$, label={lst:I_3}]
You are an expert recommender engine. You need to generate a recommendation list complying user's intention. Here is user's intention: {synthetic_intention}. Please generate a recommendation list with {item_count} different items.
Output: {item_list}

You are an expert recommender engine. You need to select a recommendation list complying user's intention from candidate items. Here is user's intention: {synthetic_intention}. The candidate items are: {candidate_titles}. Please select a recommendation list with {item_count} different items from candidate items.
Output: {item_list}
\end{lstlisting}

\begin{lstlisting}[caption=Prompts of positive intention in $I_{1}$ and $I_{3}$, label={lst:I_1_p}]
I like '{target_category}' products
Please recommend some '{target_category}' items
I'm interested in '{target_category}'
I would like to buy some '{target_category}' products
I would like to browse some '{target_category}' products
I prefer in '{target_category}' item
\end{lstlisting}

\begin{lstlisting}[caption=Prompts of negative intention in $I_{1}$ and $I_{3}$, label={lst:I_1_n}]
I don't like '{target_category}' products
Please exclude any '{target_category}' item
I'm not interested in '{target_category}'
Don't recommend me any '{target_category}' products
I don't want to browse any '{target_category}' product
I hate '{target_category}' items
\end{lstlisting}

\end{document}